\documentclass[lettersize,journal]{IEEEtran}
\usepackage{amsmath,amsfonts}
\usepackage{algorithmic}
\usepackage{algorithm}
\usepackage{array}
\usepackage[caption=false,font=normalsize,labelfont=sf,textfont=sf]{subfig}
\usepackage{textcomp}
\usepackage{stfloats}
\usepackage{url}
\usepackage{verbatim}
\usepackage{graphicx}
\usepackage{cite}
\usepackage{orcidlink}
\hypersetup{hidelinks}
\usepackage{threeparttable}
\usepackage{xcolor}
\usepackage{colortbl}
\usepackage{multirow}
\usepackage{booktabs}
\usepackage{makecell}
\usepackage{booktabs}
\usepackage{tabularx}

\usepackage{algorithm}
\usepackage{algorithmic}

\begin{document}

\title{MxGLUT: A Reconfigurable LUT-Centric Broadcast Dataflow Accelerator for Mixed-Precision GEMM}

\author{
\IEEEauthorblockN{
Weiyu~Zhou\orcidlink{0009-0002-0035-3589}\IEEEauthorrefmark{2},
\and
Chen~Ding\IEEEauthorrefmark{3},
\and
Mingyuan~Liu\IEEEauthorrefmark{3},
\and
Liangyu~Gan\IEEEauthorrefmark{2},
\and
Yukun~Feng\IEEEauthorrefmark{2},
\and
Hao~Jia\IEEEauthorrefmark{3},
\and
Haoming~Chu\IEEEauthorrefmark{3},
\and
Lirong~Zheng\IEEEauthorrefmark{3},
\and
Ning~Ma\IEEEauthorrefmark{3},
\and
Yuxiang~Huan\IEEEauthorrefmark{3}\IEEEauthorrefmark{1}
}
\\[0.8ex]
\IEEEauthorblockA{\IEEEauthorrefmark{2} University of Macau}
\IEEEauthorblockA{\IEEEauthorrefmark{3} Guangdong Institute of Intelligence Science and Technology}
}





\maketitle

\begin{abstract}
Large language model (LLM) inference suffers from growing inefficiency across the prefill and decode phases, especially under weight-only quantization, where activations remain in FP8 while weights are compressed to low-bit integers. 
Existing LUT-based accelerators mainly target FP8-INT4 computation and still rely on separate floating-point (FP) datapaths for attention GEMM operations, leading to redundant hardware and non-unified mixed-precision execution. Moreover, their static dataflows are poorly matched to the distinct prefill and decode phases.
To address these challenges, we propose MxGLUT, a reconfigurable LUT-centric broadcast (RLB) dataflow accelerator built on mixed-precision LUT-based processing elements (MxLPEs). 
Guided by a unified LUT-based execution framework, MxGLUT organizes both FP8-INT4 and FP8-FP8 GEMMs under a single LUT-based compute mechanism without dedicated FP multipliers or additional FP datapaths, and further adopts the RLB dataflow that localizes heavy partial-sum accumulation during the prefill phase and exploits weight reuse in the decode phase.
Synthesized in UMC $28\,\mathrm{nm}$ CMOS at $200~\mathrm{MHz}$, MxGLUT reduces multiplier area by up to $56.92\%$ and power by up to $77.07\%$ and $78.35\%$ in FP8-INT4 and FP8-FP8 modes, respectively. 
At the accelerator level, MxGLUT achieves an area efficiency of $0.492~\mathrm{TFLOPS/mm^2}$ and an energy efficiency of $11.58~\mathrm{TFLOPS/W}$, while adding native FP8-FP8 support incurs only $2.57\%$ and $3.34\%$ reductions in area and energy efficiency, respectively, relative to the FP8-INT4-only FIGLUT baseline. 
Across the Llama family, MxGLUT achieves up to $2.16\times$ and $1.49\times$ latency speedup, and reduces normalized energy to $0.44\times$ and $0.71\times$ in prefill and decode, respectively, with at most $1.70\%$ perplexity increase.
\end{abstract}

\begin{IEEEkeywords}
LUT, Mixed-Precision GEMM, Reconfigurable Dataflow, Accelerator.
\end{IEEEkeywords}

\section{Introduction}
\IEEEPARstart{L}{arge} language models (LLMs) have become a key workload in modern AI systems, driving growing demand for efficient inference on hardware platforms. 
Weight-only quantization has become a practical approach for reducing the memory footprint and bandwidth cost of LLM inference, where weights are compressed to low-bit integers (INT) while activations are kept in floating-point (FP) to preserve dynamic range and tolerate activation outliers~\cite{frantar2022gptq, lin2024awq, yang2024mitigating}. 
Among practical operand configurations, recent studies\cite{zhang2024qqq,lin2024qserve, luo2026posttrainingquantizationopenpangumodels} identify W4A8 as a favorable trade-off among accuracy, efficiency, and memory. 
ZeroQuant \cite{wu2023zeroquant} shows that on OPT-30B, W4A8 (FP8-INT4) has a slight increase in mean perplexity by only about 0.3\% relative to the FP16 baseline, highlighting W4A8 as a highly efficient mixed-precision operating point with negligible accuracy loss. This approach effectively reduces memory and data-transfer overhead.
However, these memory savings do not directly translate into compute efficiency, because conventional FP-INT GEMM still relies on the FP compute datapath. In FP-INT multiply-accumulate (MAC) operations, low-bit weights are typically dequantized back to FP and then processed through the same mantissa-alignment and exponent-handling logic as FP-FP MACs, leaving most FP arithmetic units active and thereby diminishing the expected energy savings~\cite{gong2024survey}. 

To address the arithmetic inefficiency of FP-INT computation in weight-only quantized inference, LUT-based acceleration has emerged as an effective approach. By precomputing activation-dependent partial products and replacing runtime multiplications with table lookups, this paradigm fundamentally reduces arithmetic complexity and power consumption in FP-INT GEMM. 
LUT-GEMM~\cite{park2022lut} precomputes partial products between low-bit weights and high-precision activations at the tile level, replacing dequantized multiplications with LUT lookups to accelerate FP-INT GEMM on GPUs. 
FIGLUT~\cite{park2025figlut} replaces traditional MAC operations with a read-accumulate (RAC) architecture, integrating a compact binary coding quantization (BCQ)~\cite{xu2018alternating}-based LUT that halves storage size. 
This design enables conflict-free parallel access, higher energy efficiency in FP-INT arithmetic, and a reduced storage footprint compared to conventional LUTs.
LUT Tensor Core~\cite{mo2025lut} proposes a hardware-software co-optimized FP-INT framework, where operator fusion and dataflow transformations reduce table-generation overhead at the software level, while a bit-serial architecture with elongated tiling enables multi-precision computation and efficient LUT reuse, offering a new architectural paradigm for low-bit inference acceleration.

Prior LUT-based RAC accelerators have demonstrated that replacing runtime FP-INT multiplications with activation-dependent table lookups can substantially improve arithmetic efficiency. However, when extended to full LLM inference, these designs remain limited in two aspects: they do not natively cover FP8-FP8 attention GEMMs, and they largely inherit static dataflows developed for inner-product MAC arrays rather than LUT-oriented outer-product execution. Moreover, these limitations correspond to two coupled forms of heterogeneity in LLM inference: cross-layer arithmetic heterogeneity across linear and attention layers, and cross-phase execution heterogeneity across prefill and decode.

At the \textbf{architecture level}, prior weight-only accelerator designs are primarily optimized for FP8-INT4 linear layers and still rely on vector processing units (VPUs)\cite{maceiras2024extending} to execute FP8-FP8 GEMMs in attention blocks. Such VPU-based execution offers limited fine-grained data reuse and relies heavily on bandwidth-intensive vector loads and stores, thereby missing the opportunity to leverage LUTs to cache intermediate results and support unified mixed-precision execution with substantially higher operand reuse. 
As quantified in Section~\ref{sec:llm}, FP8-FP8 attention GEMMs still account for a non-trivial fraction of the weighted computation cost in representative LLMs, and this fraction grows with sequence length. Therefore, an accelerator that optimizes only FP8-INT4 while offloading FP8-FP8 execution leaves a substantial portion of end-to-end inference outside its primary optimized compute fabric.
More fundamentally, LUT-based FP8-FP8 execution requires a unified execution framework compatible with existing FP8-INT4 quantization hardware, enabling seamless reuse of LUT-based datapaths with minimal additional hardware overhead, while bridging continuous FP multiplication semantics and discrete LUT indexing and jointly balancing numerical fidelity and hardware efficiency under compact representations.

At the \textbf{dataflow level}, LUT-based RAC architectures inherently organize computation as scalar-vector outer-products, a structural property fundamentally different from conventional inner-product designs~\cite{genc2021gemmini}. Yet existing LUT-based accelerators directly inherit the weight-stationary (WS) dataflow from inner-product designs without accounting for this distinction. This mismatch causes output tiles to be repeatedly shuttled between memory and compute cores, resulting in excessive output-matrix movement and energy overhead—particularly during the compute-bound prefill phase where partial-sum accumulation dominates. By contrast, the GEMV-like decode phase is characterized by memory-bound execution with repeated reuse of model weights. Therefore, a single static dataflow cannot optimize both prefill and decode, exposing the limitation of existing designs in addressing cross-phase execution heterogeneity.

To address these challenges, we propose MxGLUT, a reconfigurable LUT-centric broadcast (RLB) dataflow accelerator with mixed-precision LUT-based processing elements (MxLPEs)  for efficient LLM inference. The main contributions of this paper are as follows:
\begin{itemize}
\item We propose a unified LUT-based execution framework that maps both FP8-INT4 linear GEMMs and FP8-FP8 attention GEMMs onto the same RAC-based LUT datapath. For FP8-FP8 execution, FP product formation is moved to LUT construction, reducing runtime computation to LUT lookup and lightweight reconstruction without VPU fallback or dedicated FP datapaths.
\item We propose RLB, a LUT-centric broadcast dataflow tailored to outer-product RAC execution. By switching the stationary operand at kernel granularity, RLB keeps FP32 partial sums local in prefill and retains weights in decode, reducing partial-sum movement and avoiding a single static dataflow across heterogeneous inference phases.
\item RTL implementation in UMC $28\,\mathrm{nm}$ CMOS and end-to-end evaluation on the Llama\cite{grattafiori2024llama} family show that MxGLUT achieves unified mixed-precision execution with only marginal system-level cost while delivering substantial end-to-end efficiency gains.
\end{itemize}

\begin{figure*}[t]
    \centering
    \includegraphics[width=1\linewidth]{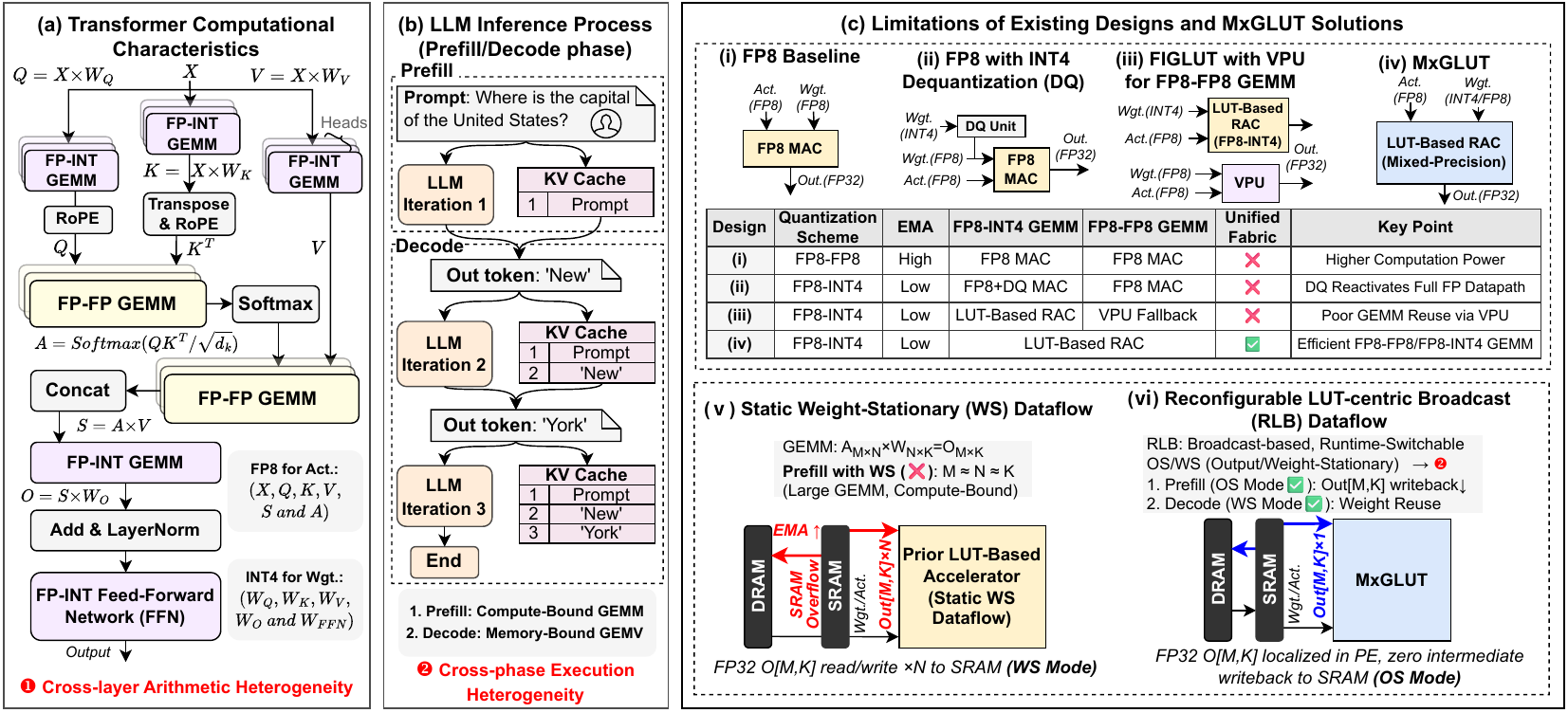}
    \caption{Overview of LLM inference heterogeneity and 
    MxGLUT design motivations: (a) transformer computational 
    characteristics with cross-layer arithmetic heterogeneity, 
    (b) prefill/decode inference process with cross-phase 
    execution heterogeneity, and (c) limitations of existing 
    designs and MxGLUT solutions.}
    \label{fig:model}
\end{figure*}

\section{Background}\label{sec:related}
\subsection{LLM Inference Characteristics}\label{sec:llm}
Transformers serve as the computational backbone of modern LLMs 
and are primarily composed of two key modules: multi-head 
self-attention (MHA) and the feed-forward network (FFN)\cite{vaswani2017attention}. Given 
an input sequence $X$, queries, keys, and values ($Q, K, V$) are 
formed by linear projections, and attention scores $A$ are computed 
from $QK^{T}$ and then applied to $V$ to produce contextual 
features $S$, as illustrated in Fig.~\ref{fig:model}(a). 
In practice, the attention path involving $QK^{T}$ and $AV$ is typically retained in FP8-FP8 GEMMs to reduce numerical error accumulation. By contrast, inference-optimized LLMs widely adopt weight-only quantization: 
projection and FFN weight matrices are stored in low-bit INT (e.g., INT4), while activations remain in FP8, processed through FP8-INT4 GEMMs. This mixed-precision scheme reduces memory footprint 
and bandwidth pressure with minimal accuracy loss. 
This \textbf{cross-layer arithmetic heterogeneity} is significant at the workload level. When computation cost is weighted by the product of operand bit-widths to reflect the effective complexity of each precision mode (i.e., FP8-FP8 operations carry twice the weight of FP8-INT4 operations), FP8-FP8 attention GEMMs still account for $\sim$52\% of the total weighted computation cost in Llama-3-1B during 8K-token prefill, and this fraction further increases with sequence length due to the quadratic scaling of attention.

As illustrated in Fig.~\ref{fig:model}(b), LLM inference comprises two structurally distinct phases: prefill and decode, each exhibiting fundamentally different computational characteristics—a phenomenon we refer to as \textbf{cross-phase execution heterogeneity}.
During the prefill phase, the model processes all input 
prompt tokens in parallel, generating the first output token while 
simultaneously computing and caching the key-value (KV) pairs 
across all transformer layers. Because the entire prompt is processed as a single large matrix operation, the prefill phase is inherently compute-bound, dominated by large matrix-matrix GEMMs that demand extensive data reuse to alleviate memory access overhead.
During the decode phase, tokens are generated autoregressively one at a time, with each step taking the most recently generated token and the accumulated 
KV-cache as inputs. 
Due to the single-token generation pattern, computation degenerates into a memory-bound GEMV-like operation, where each output is produced through repeated application of model weights over the accumulated sequence dimension. This results in limited arithmetic intensity and pronounced weight reuse across decoding steps, shifting the execution bottleneck toward memory bandwidth. 
The two phases therefore exhibit fundamentally different compute and memory access characteristics, posing significant challenges for dataflow scheduling in LLM inference accelerators.

\subsection{LLM Accelerator and Dataflow Designs}  
Early transformer accelerators~\cite{tambe202216, kim202420, 10631541} primarily targeted FP16 or INT8 matrix operations through systolic arrays and structured sparsity. However, as weight-only quantization emerged as a mainstream deployment strategy, conventional FP datapaths became increasingly inefficient for FP-INT computation, since dequantization reactivates the full FP multiply path and negates the expected energy savings. This has motivated LUT-based acceleration, which eliminates runtime FP-INT multiplications by precomputing activation-dependent partial products as lookup table entries.
LUT-GEMM~\cite{park2022lut} first demonstrated this paradigm on GPUs by precomputing partial products between low-bit weights and high-precision activations at the tile level, replacing dequantized multiplications with table lookups. FIGLUT~\cite{park2025figlut} brought this idea to dedicated hardware by replacing conventional MAC units with a RAC datapath, employing BCQ to halve LUT storage and enabling conflict-free parallel access with improved FP-INT energy efficiency. LUT Tensor Core ~\cite{mo2025lut} further co-optimizes hardware and software through operator fusion and dataflow transformations to reduce table-generation overhead, adopting a bit-serial architecture with elongated tiling for multi-precision computation and aggressive LUT reuse. Yet these designs are developed exclusively for FP-INT computation, with FP-FP attention GEMMs remaining outside the LUT-based execution framework.

Dataflow design plays an equally critical role in determining the efficiency of GEMM accelerators. The WS and OS dataflows, formalized in works such as Gemmini~\cite{chen2014dadiannao,jouppi2017datacenter,genc2021gemmini}, represent the dominant paradigms for controlling data reuse and partial-sum movement in systolic arrays. Beyond these canonical dataflows, INCA~\cite{kim2023inca} demonstrates that input-stationary dataflow can fundamentally reduce memory access overhead in processing-in-memory (PIM) architectures by retaining activations in compute arrays rather than shuttling them through external buffers. 
DPIMA~\cite{kim2025dpima}, a DRAM-based PIM accelerator for privacy-preserving machine learning, further shows that different computing phases may exhibit inherently heterogeneous characteristics, thereby motivating phase-aware dataflow switching rather than a single static mapping. However, its design is specialized for DRAM-PIM architectures and does not address the partial-sum movement overhead that arises in LUT-based outer-product execution.
Unlike conventional inner-product designs, LUT-based RAC architectures inherently organize computation as scalar-vector outer-products. Among existing designs, FIGLUT~\cite{park2025figlut} adopts a WS dataflow to maximize weight reuse, while LUT Tensor Core~\cite{mo2025lut} is designed as a drop-in replacement for GPU Tensor Cores, focusing on software-hardware co-optimization for LUT generation overhead reduction within the existing GPU execution framework.

\subsection{Limitations and Motivation}
As illustrated in Fig.~\ref{fig:model}(c), existing designs exhibit two fundamental limitations that hinder efficient LLM inference.
At the \textbf{architecture level}, conventional FP8 designs store both weights and activations in FP8 format, resulting in higher external memory access (EMA) due to the larger weight storage footprint compared with low-bit alternatives. Although FP8-INT4 dequantization-based designs reduce EMA by compressing weights to INT4, dequantization reactivates the full FP multiply path at runtime, negating much of the expected energy reduction.
While LUT-based designs such as FIGLUT successfully eliminate FP8-INT4 multipliers and reduce both EMA and power consumption, they still rely on VPUs to handle FP8-FP8 GEMMs in attention layers. Such VPU-based execution offers limited fine-grained data reuse and depends heavily on bandwidth-intensive vector loads and stores, leaving FP8-FP8 attention computation outside the LUT-based execution path.
Extending LUT-based execution to FP8-FP8 is non-trivial due to the mismatch between continuous FP multiplication and discrete LUT indexing. This motivates a unified execution framework that directly reuses the FP8-INT4 LUT infrastructure to support FP8-FP8 GEMMs with minimal additional hardware overhead while preserving numerical fidelity.

At the \textbf{dataflow level}, as shown in Fig.~\ref{fig:model}(c)(v),
existing LLM accelerators adopt a static WS dataflow inherited
from conventional inner-product designs. However, unlike
inner-product designs where activations and weights are consumed
symmetrically, LUT-based RAC architectures first construct
activation-dependent LUT entries and then broadcast them across
multiple weights in parallel, inherently organizing computation
as scalar-vector outer-products. Each step computes an
$(M \times 1) \times (1 \times K)$ outer-product and must be
repeated $N$ times to complete a full GEMM of
$\mathbf{A}_{M \times N} \times \mathbf{W}_{N \times K}
= \mathbf{O}_{M \times K}$. This structural property forces the
output matrix $\mathbf{O}[M, K]$ to be read and written back to
on-chip SRAM $N$ times throughout execution. During the
compute-bound prefill phase where $M \approx N \approx K$,
this repeated writeback rapidly exhausts on-chip SRAM capacity,
triggering SRAM overflow and substantially increasing EMA. 
Meanwhile, in the memory-bound decode phase, WS can achieve higher weight reuse than OS in GEMV-like decode operations, as weights are repeatedly accessed across multiple input activations.
Overall, a single static WS dataflow cannot simultaneously address the heavy partial-sum writeback pressure in the prefill phase and the GEMV-like execution pattern in the decode phase, highlighting the need for a more flexible dataflow to handle the cross-phase execution heterogeneity inherent in LLM inference.

To address these two limitations jointly, we propose MxGLUT, as illustrated in Fig.~\ref{fig:model}(c)(iv)(vi). MxGLUT introduces a unified LUT-based execution framework for FP8-INT4 and FP8-FP8 GEMMs, together with the RLB dataflow that enables runtime switching between OS and WS execution modes: OS localizes partial-sum accumulation during prefill to eliminate intermediate writeback to SRAM, while WS leverages weight reuse during decode to alleviate memory bandwidth pressure.

\section{LUT-based Mixed-Precision GEMM}\label{sec:methodology}
This section presents the proposed LUT-based mixed-precision GEMM from two perspectives. Section~\ref{sec:abstraction} first introduces a unified LUT-based execution framework for both FP8-INT4 and FP8-FP8 GEMMs. Section~\ref{sec:mxlpe} then presents the MxLPE microarchitecture, which implements this framework in hardware.

\subsection{Unified LUT-based Execution Framework}
\label{sec:abstraction}
MxGLUT targets weight-only LLM inference, where FP8-INT4 linear layers and FP8-FP8 attention layers need to run on the same LUT-centric compute fabric. The proposed unified LUT-based execution framework maps both GEMM types onto a common runtime path, avoiding separate arithmetic engines or dedicated FP datapaths. This is challenging because FP8-FP8 execution, unlike FP8-INT4 computation with low-bit encoded weights, must still handle FP multiplication steps such as mantissa multiplication, normalization, rounding, sign processing, and exponent-bias correction within a LUT-based design. 

Under the unified LUT-based execution framework, MxGLUT addresses this mismatch by organizing both modes into the same three-stage execution flow:
(i) \textit{LUT construction}, where activation-dependent partial results are precomputed and stored as LUT entries;
(ii) \textit{LUT query}, where encoded weights are used to index the corresponding LUT entries;
and (iii) \textit{mode-dependent transformation}, where the queried results are converted into final partial results according to the target precision mode.
This shared execution pattern allows the central LUT query path to be reused across heterogeneous mixed-precision GEMM operations, while confining precision-dependent logic to LUT construction and lightweight post-query transformation.
The following subsections instantiate this framework for FP8-INT4 and FP8-FP8 GEMMs, respectively, highlighting how their distinct numerical requirements are accommodated within the same LUT-centric execution flow.


\begin{figure}[!t]
    \centering
    \includegraphics[width=1\linewidth]{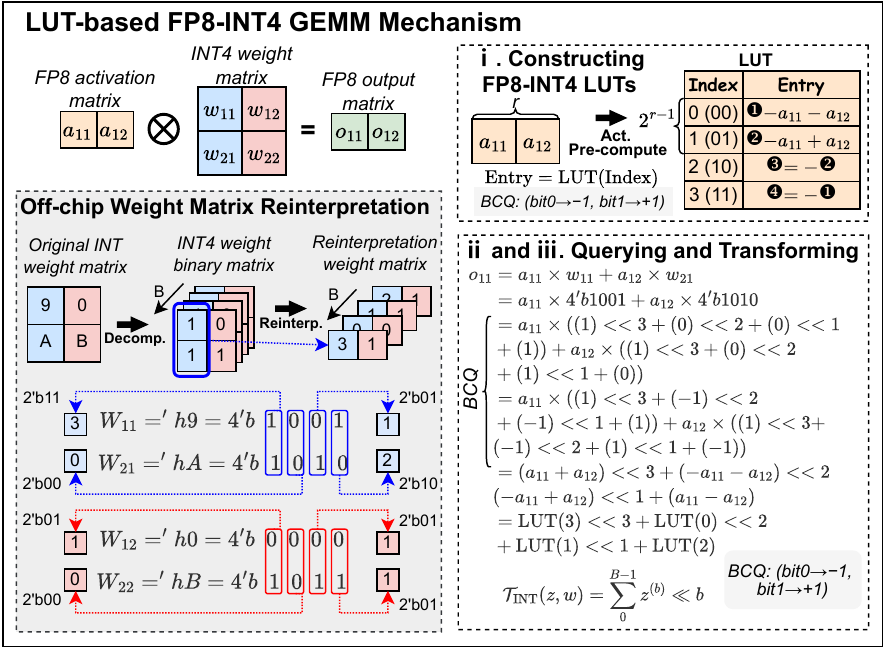}
    \caption{Instantiation of the unified LUT-based execution framework for FP8-INT4 GEMM.
    }
    \label{fig:int4_method}
\end{figure}

\subsubsection{LUT-based FP8-INT4 GEMM}
Under the LUT-based execution framework, the FP8-INT4 mode is implemented through LUT indexing with low-bit encoded weights followed by shift-based accumulation of the queried results.
In this formulation, the encoded weight index is derived through an off-chip preprocessing step: low-bit INT weights are decomposed into bit planes and regrouped by columns to form a reinterpretation weight matrix, where each element is represented as a bit-packed column-wise binary code obtained by stacking all bits in the same column into a new INT. This encoded representation later serves as part of the LUT index. 

As illustrated in Fig.~\ref{fig:int4_method}, the overall procedure can be described as follows. 
(\romannumeral1) \textit{LUT construction}, where activations are grouped into vectors of size $r$, and BCQ sign patterns (bit~0 $\to -1$ and bit~1 $\to +1$) are used to enumerate activation combinations. By exploiting sign symmetry, only $2^{r-1}$ representative patterns need to be stored, thereby halving the LUT storage while precomputing the corresponding FP8 combinations as LUT entries;
(\romannumeral2~and~\romannumeral3) \textit{LUT query and mode-dependent transformation}, where the bit-packed column-wise weight code is used to index the LUT during inference, and the queried FP8 combinations are further shift-accumulated across bit planes to form the final partial results.
Here, $r=2$ is used for illustration, while $r=4$ is adopted in the hardware implementation to match the baseline configuration in prior LUT-based designs~\cite{park2025figlut}.


\begin{algorithm}[t]
\caption{FP8$\times$FP8 LUT-based GEMM}
\label{alg:fp8fp8gemm}
\begin{algorithmic}[1]
 
\REQUIRE FP8 activation matrix $\mathbf{A}_{M \times N}$;
         FP8 weight matrix $\mathbf{W}_{N \times K}$
\ENSURE  FP32 output matrix $\mathbf{O}_{M \times K}$

\FOR{$i \gets 0$ \TO $M\!-\!1$}
  \FOR{$n \gets 0$ \TO $N\!-\!1$}

    \STATE $\{s_a,\; e_a,\; m_a\} \gets \mathbf{A}_{i,\,n}$
           \COMMENT{decompose FP8 activation}
    \STATE $M_a \gets \{1,\; m_a\}$
           \COMMENT{prepend implicit leading 1}
    \STATE $s_{\rm LUT} \gets s_a$
        \COMMENT{stored as separate sign metadata}
    \FOR{$q \gets 0$ \TO $7$}
      \STATE $M_q \gets \{1,\; q\}$
             \COMMENT{enumerate all weight mantissa candidates (4-bit)}
      \STATE $P \gets M_a \times M_q$
      \STATE $\delta \gets P[7]$
             \COMMENT{overflow bit}
      \STATE $m_{\rm n} \gets \mathrm{RoundNorm}(P)$
      \STATE $E_{\rm LUT} \gets \mathrm{SExt}(e_a) + \delta - Bias$
             \COMMENT{5-bit signed intermediate exponent}
      \STATE $\mathrm{LUT}_q \gets \{E_{\rm LUT}[4:0],\; m_{\rm n}[2:0]\}$

    \ENDFOR

    \FOR{$k \gets 0$ \TO $K\!-\!1$}
      \STATE $\{s_w,\; e_w,\; m_w\} \gets \mathbf{W}_{n,\,k}$
      \STATE $\{E_z,\; m_z\} \gets \mathrm{LUT}_{m_w}$
      \STATE $s_z \gets s_{\rm LUT} \oplus s_w$
      \STATE $E_p \gets E_z + \mathrm{SExt}(e_w)$
      \STATE $\mathbf{O}_{i,\,k} \mathrel{+}=
             \{s_z,\; E_p,\; m_z\}$
             \COMMENT{no multiplier at runtime}
    \ENDFOR
  \ENDFOR
\ENDFOR
\RETURN $\mathbf{O}$
\end{algorithmic}
\end{algorithm}

\subsubsection{LUT-based FP8-FP8 GEMM}
FP8-FP8 operations are not naturally compatible with LUT-based execution using bit-plane decomposition and shift-based accumulation, due to the mismatch between continuous FP semantics and discrete LUT indexing.
To bridge this mismatch, MxGLUT maps FP8-FP8 computation to a LUT-centric execution view, where FP product formation is moved from runtime computation to LUT construction. Specifically, mantissa multiplication, normalization, and round-to-nearest (RTN) are performed when generating LUT entries, while runtime execution is reduced to LUT indexing, sign combination, exponent addition, and FP32 accumulation. 
Subnormal values are eliminated using flush-to-zero (FTZ), and each product is represented by a compact 8-bit FP8-like LUT entry rather than an FP16/FP32 product, reducing LUT storage and simplifying post-query reconstruction. 
The end-to-end accuracy impact of FTZ and the compact FP8-like product representation is evaluated in Section~\ref{sec:ppl}.

\begin{figure}[!t]
    \centering
    \includegraphics[width=1\linewidth]{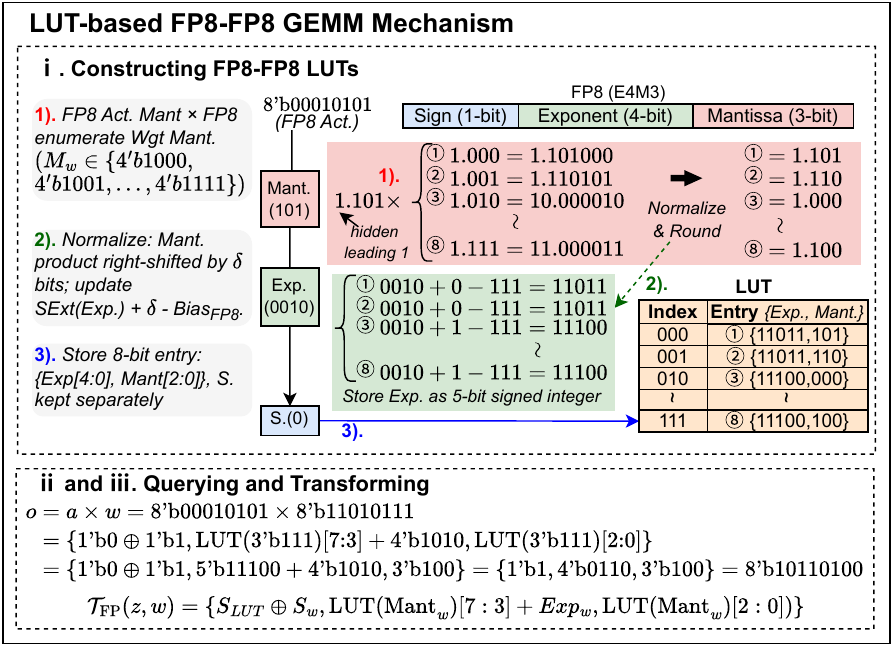}
    \caption{Instantiation of the unified LUT-based execution framework for the proposed FP8-FP8 GEMM.
    }
    \label{fig:fp8_method}
\end{figure}

As illustrated in Fig.~\ref{fig:fp8_method} and detailed in Lines 6$\sim$13 of Algorithm~\ref{alg:fp8fp8gemm}, the proposed FP8-FP8 mechanism follows the same three-stage flow:
(\romannumeral1) \textit{LUT construction}, where all possible FP8 weight mantissas are enumerated, multiplied with activation mantissas, normalized, rounded, and packed into LUT entries containing a 5-bit signed exponent-related field and a 3-bit mantissa-related field;
(\romannumeral2~and~\romannumeral3) \textit{LUT query and mode-dependent transformation}, where each encoded FP8 weight uses its mantissa bits to index LUT entries during inference, and the queried results are combined with lightweight sign XOR and exponent addition to reconstruct the final partial results. Here, $\mathrm{SExt}(\cdot)$ denotes hardware sign extension, which expands the FP8 exponent to the accumulator exponent width before addition.

\begin{figure*}[!t]
    \centering
    \includegraphics[width=1\linewidth]{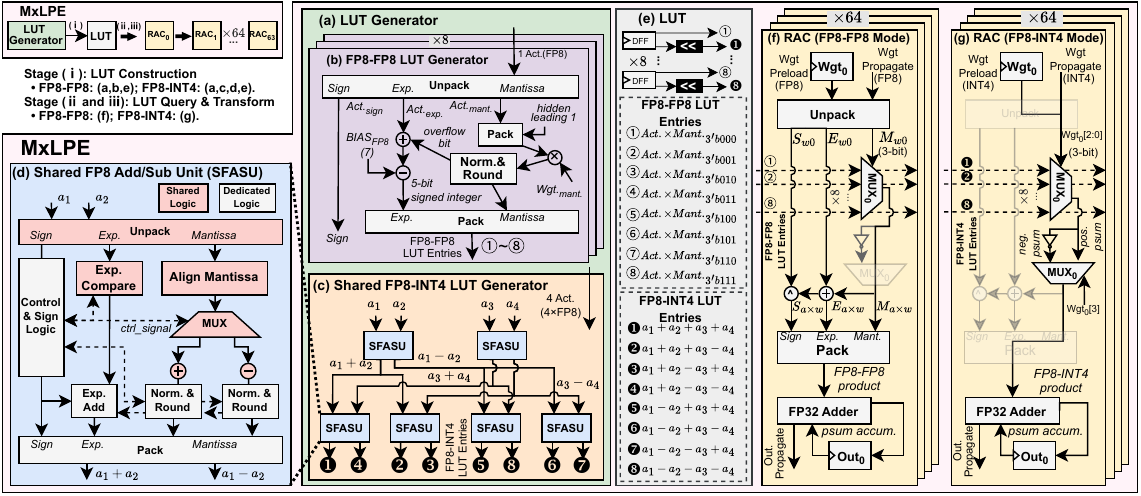}
    \caption{
    Microarchitecture of the mixed-precision LUT-based processing element (MxLPE), (a) LUT generator, (b) FP8-FP8 LUT generator, (c) shared FP8-INT4 LUT generator, (d) shared FP8 add/sub unit (SFASU), and (e) LUT; datapaths of the read-accumulate (RAC) unit for (f) FP8-FP8 and (g) FP8-INT4 modes.
    }
    \label{fig:microarch}
\end{figure*}

\subsection{Mixed-Precision LUT-based Processing Element (MxLPE)}\label{sec:mxlpe}

The MxLPE implements the LUT-based execution framework through a LUT-centric datapath supporting both FP8-INT4 and FP8-FP8 modes. As illustrated in the upper-left part of Fig.~\ref{fig:microarch}, it integrates a LUT generator, a local LUT block, and multiple parallel RAC units.
The hardware execution path follows the same order as the LUT-based execution framework. 
In stage~(\romannumeral1), \textit{LUT construction}, input operands are first preprocessed by the LUT generator to produce activation-dependent LUT entries, which are then stored in the local LUT block. In stage~(\romannumeral2~and~\romannumeral3), \textit{LUT query and transformation}, encoded weights index the local LUT entries, and the selected values are forwarded to RAC units for lightweight transformation and accumulation. 
In this way, MxLPE avoids runtime multiplication and performs inference through LUT query, lightweight transformation, and FP32 accumulation.

\subsubsection{LUT Generator and LUT Block Design}
As shown in Fig.~\ref{fig:microarch}(a)(b)(c)(d)(e), the LUT subsystem consists of a dedicated FP8-FP8 LUT generator, a shared FP8-INT4 LUT generator, and a flip-flop–based LUT block that stores the generated entries.
In stage~(\romannumeral1), \textit{LUT construction}, the FP8-FP8 and FP8-INT4 paths follow Fig.~\ref{fig:microarch}(a)(b)(e) and Fig.~\ref{fig:microarch}(a)(c)(d)(e), respectively, to generate their corresponding LUT entries.

As shown in Fig.~\ref{fig:microarch}(b), the FP8-FP8 LUT generator performs mantissa enumeration and pre-normalization during LUT construction using small fixed-point arithmetic blocks.
When a single FP8 activation is fetched from the activation SRAM, the unpack unit decomposes it into a 1-bit sign, a 4-bit exponent, and a 3-bit mantissa. The mantissa is extended with the implicit leading 1 to form a 4-bit mantissa, which is then multiplied by each enumerated 4-bit weight mantissa with the implicit bit in the range $4'b1000 \sim 4'b1111$. The product is normalized by the round-and-shift block, which preserves a 4-bit mantissa (including the implicit bit) and records the shift amount; this shift amount is added to the 4-bit exponent.
Finally, the updated exponent is stored as a 5-bit signed intermediate exponent, and the rounded 3-bit mantissa is packed with it into an 8-bit LUT entry, while the activation sign is kept as separate sign metadata.

The shared FP8-INT4 LUT generator mainly consists of six SFASUs, as shown in Fig.~\ref{fig:microarch}(c)(d). When four FP8 activations ($a_1,a_2,a_3,a_4$) are fetched from the activation SRAM, BCQ encodes them into bipolar values (bit~0 $\rightarrow -1$, bit~1 $\rightarrow +1$), which are then combined to construct the LUT input vector. By exploiting sign symmetry among LUT entries, e.g., $a_1+a_2+a_3+a_4=-(-a_1-a_2-a_3-a_4)$, the generator enumerates the eight unique sign combinations and precomputes the corresponding FP8 sums as LUT entries, as illustrated in Fig.~\ref{fig:microarch}(e).
By sharing operand unpacking and exponent alignment across add/sub paths, only the final sign-dependent accumulation is executed in parallel, reducing duplicated preprocessing compared with prior designs~\cite{park2025figlut}.

The LUT block consists of eight flip-flop–based LUTs, each 8 bits wide, together with eight per-entry FP shifters. These FP shifters are dedicated to bit-plane output recombination in the FP8-INT4 mode. 
In a conventional design~\cite{mo2025lut}, each RAC instantiates a dedicated FP shifter tightly coupled with its FP32 adder to perform shift-accumulate operations. In contrast, our design broadcasts each LUT entry to all parallel RACs after performing FP shifting once per entry, so that shifting is shared across RACs rather than replicated per RAC. This eliminates redundant FP shifters, substantially reducing both area and energy consumption.

\subsubsection{Read-Accumulate (RAC) Unit Design}

As shown in Figs.~\ref{fig:microarch}(f)(g), the RAC unit serves as the unified execution backbone of MxGLUT. Across both precision modes, each RAC selects a LUT-derived value, performs lightweight product reconstruction, and accumulates the result into an FP32 partial sum.
Each RAC contains LUT-selection MUXes, lightweight reconstruction logic, an FP32 adder, and local weight/output registers that support both WS and OS dataflows.
In stage~(\romannumeral2~and~\romannumeral3), \textit{LUT query and transformation}, both modes read entries from the local LUT block, but use different transformation paths.
FP8-FP8 follows Fig.~\ref{fig:microarch}(f), where the selected LUT entry is combined with the activation sign and the original weight sign/exponent before accumulation.
FP8-INT4 follows Fig.~\ref{fig:microarch}(g), where the selected LUT entries are recombined according to the encoded low-bit weight.

In FP8-FP8 mode, the unpack unit decomposes the FP8 weight into sign, exponent, and mantissa fields. The FP8-FP8 LUT generator stores the precomputed LUT entries in the LUT block, and the 8-to-1 multiplexer (MUX) selects one entry according to the 3-bit mantissa field of the weight. In this mode, the 2-to-1 MUX is bypassed. 
Note that the LUT block fans out to the 8-to-1 MUX of every RAC, so each LUT entry can be shared across all RAC units. 
The queried LUT entry is then combined with the activation sign and the original weight sign/exponent, where sign XOR produces the product sign and exponent addition updates the product exponent.
The reconstructed sign, exponent, and mantissa are then packed back into an FP8-like product and aligned to the FP32 adder for accumulation. 

In FP8-INT4 mode, the shared FP8-INT4 LUT generator stores the precomputed LUT entries into the LUT block. The 8-to-1 MUX selects one of the eight entries according to the lower three bits of the INT4 weight, while the 2-to-1 MUX uses the most significant bit to choose between the selected value and its negation. This leverages LUT-entry sign symmetry so that only eight entries need to be stored. The selected value is then directly forwarded to the FP32 adder to update the running partial sum. 
The per-entry FP shifters in the LUT block, controlled by a local state machine, pre-shift the LUT entries associated with different bit-planes prior to accumulation, enabling efficient bit-plane recombination in FP8-INT4 mode. As a result, both FP8-INT4 and FP8-FP8 execution share the same accumulation backbone, while differing only in the LUT-selection and lightweight reconstruction steps before FP32 accumulation.

\begin{figure}[t]
    \centering
    \includegraphics[width=0.95\linewidth]{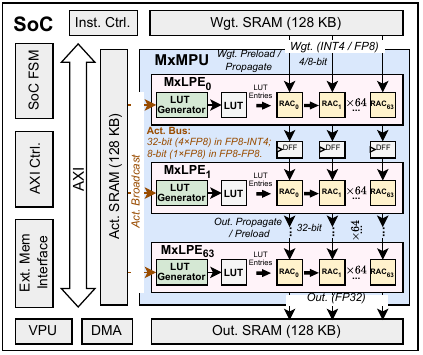}
    \caption{Overall architecture of MxGLUT.}
    \label{fig:overall_arch}
\end{figure}

\section{Proposed Architecture}
\label{sec:arch}
\subsection{Architecture Overview}
\label{sec:overview_arch}

The overall architecture of the proposed accelerator is illustrated in Fig.~\ref{fig:overall_arch}. 
The accelerator comprises a mixed-precision MPU (MxMPU), a 256-bit AXI interconnect with its control unit, a DMA engine, an instruction controller, an on-chip SRAM subsystem, and SoC-level control logic, interfacing with off-chip DRAM via a standard AXI bus and relying on the hierarchical on-chip SRAM to buffer activations, weights, and intermediate results.
The VPU is retained only for nonlinear operators such as Softmax and LayerNorm and is not involved in GEMM execution in MxGLUT.
All FP8-INT4 and FP8-FP8 GEMMs are executed natively in the MxMPU.
The subsystem integrates 16 activation SRAM macros, 16 weight SRAM macros, and 16 output SRAM macros. Each macro has a capacity of 8 KB; the activation and output SRAMs use 128-bit interfaces, while the weight SRAMs use 32-bit interfaces. In total, the subsystem provides 384 KB of on-chip storage. Each SRAM adopts a two-bank organization to support ping-pong double buffering between the compute datapath and the DMA engine, enabling overlapped computation and data transfer.

At the computation core, the MxMPU implements a $64 \times 64$ RAC compute fabric organized as 64 MxLPEs. The activation bus operates in two configurations: 32-bit mode (carrying $4\times$ FP8 activations) for FP8-INT4 execution, and 8-bit mode (carrying $1\times$FP8 activation) for FP8-FP8 execution. Correspondingly, the weight path delivers INT4 weights in FP8-INT4 mode and FP8 weights in FP8-FP8 mode.
By leveraging the RLB dataflow, the MxMPU adapts its data-propagation direction and scheduling to the current LLM inference phase (prefill or decode) and computation mode (FP8-INT4 or FP8-FP8), thereby sustaining high efficiency under heterogeneous workloads. 

\begin{figure*}[t]
    \centering
    \includegraphics[width=0.98\linewidth]{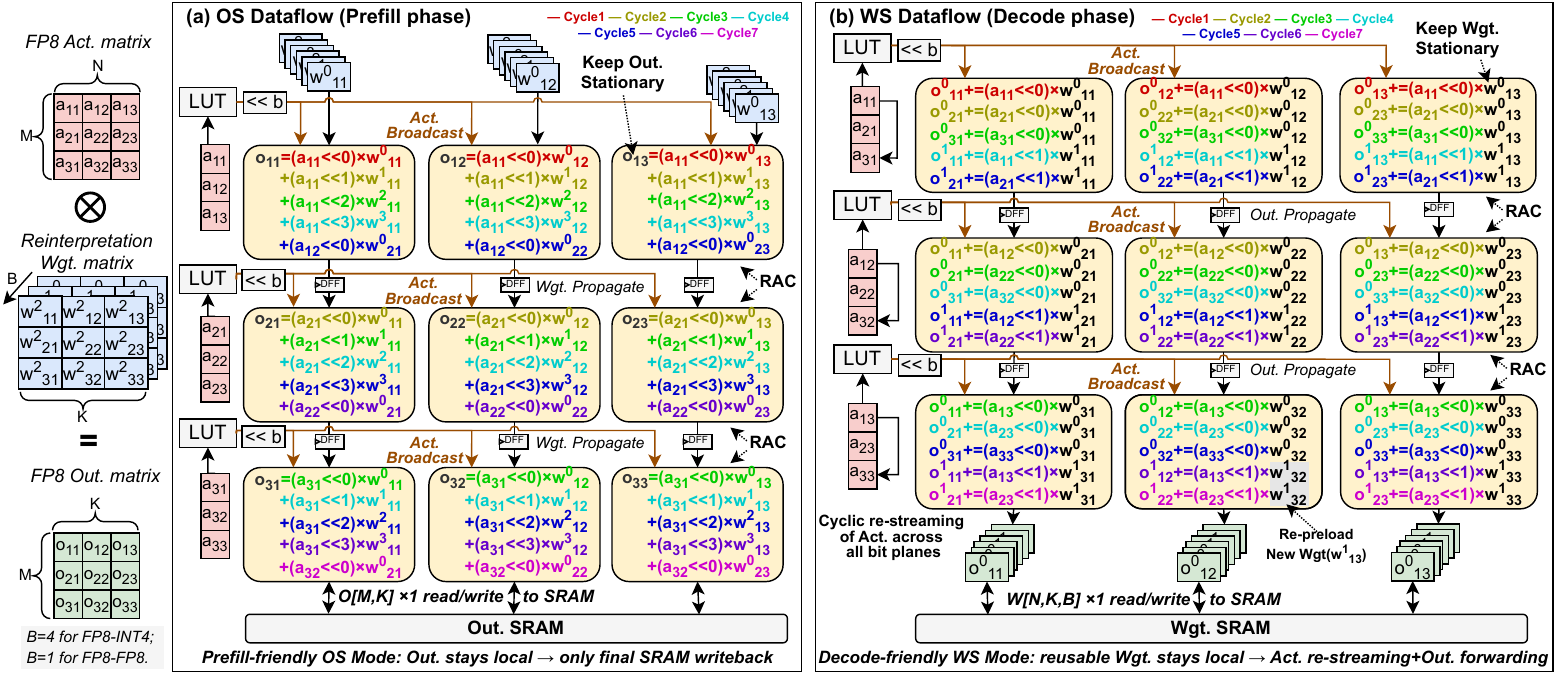}
    \caption{ 
    Illustration of the proposed reconfigurable LUT-centric broadcast (RLB) dataflow operating in (a) output-stationary (OS) mode and (b) weight-stationary (WS) mode.
    }   
    \label{fig:dataflow}
\end{figure*}

\subsection{Reconfigurable LUT-Centric Broadcast (RLB) Dataflow}\label{sec:dataflow}

MxGLUT adopts the RLB dataflow, which performs row-wise broadcast of activation-derived LUT entries and supports array-level switching between OS and WS mappings, with only lightweight control reconfiguration and negligible runtime overhead. Unlike conventional systolic dataflows that propagate activation-side operands across neighboring PEs, RLB directly delivers LUT entries to multiple RACs in parallel. As a result, wavefront fill/drain overhead is removed, reducing execution latency from $3N+S-3$ to $2N+S-2$, and the data-distribution register overhead is reduced from $\frac{3N(N-1)}{2}$ to $\frac{N(N-1)}{2}$, as summarized in Table~\ref{tab:dataflow}.
With propagation overhead significantly reduced by row-wise LUT broadcast, the main performance bottleneck becomes on-chip partial-sum storage and buffering. Since LUT-based GEMM in MxGLUT is organized as scalar-vector outer-product execution, each broadcasted LUT batch contributes to multiple output elements within a row of RACs. As a result, dataflow behavior is primarily determined by which operands are kept stationary, i.e., partial sums or weights during execution.

\begin{table}[!t]
    \centering
    \caption{Comparison of latency, throughput, and register overhead under different dataflows}
    \label{tab:dataflow}
    \begin{threeparttable}
    \begin{tabular}{cccc}
        \toprule
        \textbf{Dataflows} & \textbf{Latency} & \textbf{Throughput} & \textbf{Overhead}$^{\dagger}$ \\
        \midrule
        Systolic WS  & \multirow{2}{*}{$3N+S-3$} & \multirow{2}{*}{$\frac{2N^3}{3N+S-3}$} & \multirow{2}{*}{$N(N-1)$}\\
        or OS Only\cite{abdelmaksoud2025dip} &  & & \\
        \midrule
        Reconfig. & \multirow{2}{*}{$3N+S-3$} & \multirow{2}{*}{$\frac{2N^3}{3N+S-3}$} & \multirow{2}{*}{$\frac{3N(N-1)}{2}$}\\
        Systolic WS + OS &  & & \\
        \midrule
        RLB & $2N+S-2$ & $\frac{2N^3}{2N+S-2}$ & $\frac{N(N-1)}{2}$\\
        \bottomrule
    \end{tabular}
    \end{threeparttable}
    \begin{threeparttable}
        \begin{tablenotes}    
        \footnotesize
        \item[$\ddag$] $N$ denotes the number of rows/columns in the systolic array, and $S$ denotes the MAC pipeline depth. 
        \item[$\dagger$] Overhead denotes the data-distribution register overhead required for operand delivery and synchronization across the PE array.
      \end{tablenotes} 
    \end{threeparttable}
\end{table}

As illustrated in Fig.~\ref{fig:dataflow}, the operation of the RLB dataflow in FP8-INT4 mode is explained using an illustrative $3\times3$ output tile mapped onto three rows of MxLPEs, each containing three RAC units.
In the OS dataflow shown in Fig.~\ref{fig:dataflow}(a), each RAC is assigned to one output element $o_{ik}$, which remains stationary in the RAC while all partial contributions are accumulated locally. 
For a given activation group, e.g., $a_{11}$ in the FP8-INT4 example, the activation bus delivers $4\times$FP8 values through a 32-bit bus. In contrast, the same bus interface is configured to deliver $1\times$FP8 activation through an 8-bit bus in FP8-FP8 mode.
The activations then enter the LUT generator to produce eight LUT entries. After bit-plane shifting ($\ll b$), where $b \in \{0,\dots,B-1\}$ and $B=4$ in FP8-INT4 mode (while $B=1$ in FP8-FP8 mode), these LUT entries are broadcast to all RACs in the same row, while the corresponding weight bit-planes $w^b_{jk}$ propagate horizontally across the row.
Specifically, in FP8-INT4 mode, the bit-plane shift is handled by the LUT block, so that shifted LUT entries are produced in advance and no dedicated shift logic is required in each RAC. 
Each RAC therefore selects a pre-shifted LUT entry and accumulates it into the local FP32 accumulator associated with $o_{ik}$. The different box colors indicate successive cycles. Across multiple cycles and bit planes, all contributions to $o_{ik}$ are accumulated in place, allowing the final FP32 output matrix to be generated without writing intermediate partial sums back to SRAM.

In the WS dataflow of Fig.~\ref{fig:dataflow}(b), the stationarity pattern of weights and outputs is interchanged. Each RAC keeps its weight $w^b_{jk}$ stationary, while output elements are propagated across RACs and accumulated through output forwarding.
Activations are cyclically re-streamed across all bit planes through the LUT generator, while the stationary weights are successively applied across these bit planes, enabling each weight to participate in multiple outer-product updates. This bit-plane–wise reuse allows the contributions from all bit planes to a given output tile to be merged within a single pass through the array, reducing on-chip partial-sum buffering and shifter activity. 
At the same time, the WS dataflow improves decode-phase efficiency by promoting weight reuse in GEMV-like execution patterns, thereby reducing memory bandwidth demand associated with weight access in autoregressive inference.

Accordingly, RLB selects OS when the dominant cost of a workload is partial-sum accumulation and buffering, and selects WS when it is dominated by weight reuse. In LLM inference, these two regimes naturally correspond to the prefill and decode phases, respectively. The OS/WS mode is selected at array granularity for each kernel invocation; thus, mode switching only changes the placement of stationary operands and the data propagation control, with negligible overhead compared to GEMM execution.

\begin{figure*}[!t]
    \centering
    \includegraphics[width=0.99\linewidth]{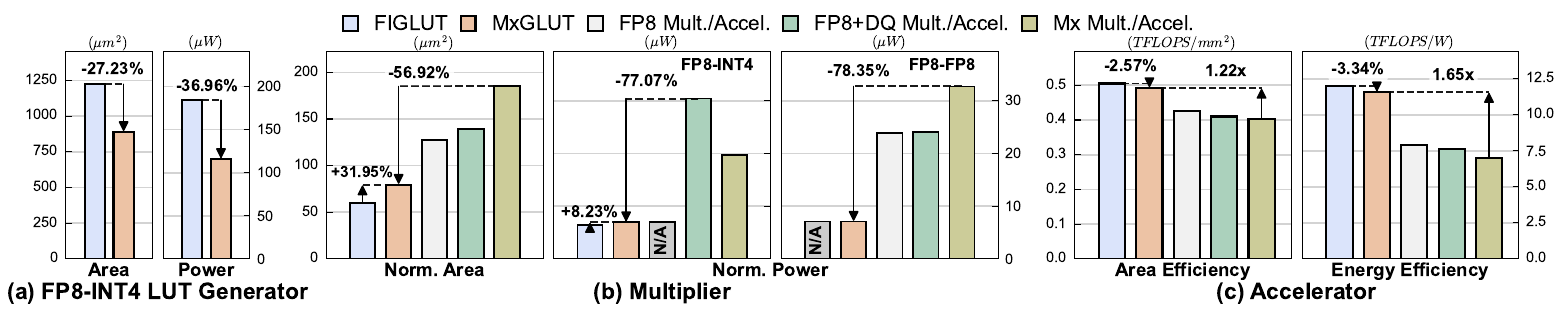}
    \caption{Architectural evaluation results. (a) Area and power comparison between the proposed shared FP8-INT4 LUT generator and the conventional design~\cite{park2025figlut}. (b) Normalized area and power comparison of the proposed MxGLUT multiplier against baseline multipliers under FP8-INT4 and FP8-FP8 modes. (c) Area and energy efficiency comparison between MxGLUT and baseline accelerators.}
    \label{fig:mult}
\end{figure*}

\section{Experimental Results} \label{sec:res}
\subsection{Experimental Setup}
The proposed MxGLUT, together with the baseline designs, is fully implemented in Verilog and verified using Synopsys VCS. 
Logic synthesis uses Synopsys Design Compiler under a $200~\mathrm{MHz}$ timing constraint (UMC $28\,\mathrm{nm}$ library), and power is estimated with Synopsys PrimeTime PX. 
Both on-chip SRAM buffers and off-chip DRAM energy are modeled using CACTI tool\cite{balasubramonian2017cacti}.
To ensure a fair architectural comparison, all evaluated accelerators use the same clock target, array dimension, on-chip SRAM budget, AXI/DMA subsystem, and top-level memory interface. For each architecture, we apply the same Timeloop-based~\cite{parashar2019timeloop} optimization flow and report results under its own best legal mapping. Therefore, the reported differences primarily reflect the effects of compute-path design and dataflow organization, rather than mismatched system budgets or suboptimal scheduling.

\subsection{Architectural Evaluation}
\subsubsection{Arithmetic-Level Evaluation}
We compare the proposed shared FP8-INT4 LUT generator with a conventional FP8-INT4 LUT generator~\cite{park2025figlut} under identical technology and configuration settings. 
As shown in Fig.~\ref{fig:mult}(a), the proposed generator reduces silicon area by $27.23\%$ and power by $36.96\%$ , demonstrating the effectiveness of the shared FP8 front-end datapath in eliminating redundant add/sub preprocessing logic.

\subsubsection{PE-Level Evaluation} 
We evaluate the proposed MxGLUT multiplier (i.e., the MxLPE datapath excluding the FP32 adder) against four baseline designs under FP8-INT4 and FP8-FP8 modes in terms of normalized area and power.
All results are normalized to equivalent compute throughput for fair comparison. 
The baselines represent four typical design choices: 
a native FP8-FP8 multiplier (FP8 Mult.); an FP8 datapath augmented with INT4 dequantization (FP8+DQ Mult.); a dual-path mixed-precision design (Mx Mult.) with separate FP8-FP8 and FP8-INT4 units, where FP8-INT4 multiplication is implemented using a bit-serial shift-and-add scheme~\cite{kim2023winning}; and FIGLUT as a LUT-only FP8-INT4 baseline.
As shown in Fig.~\ref{fig:mult}(b), MxGLUT achieves up to $56.92\%$ area reduction and up to $77.07\%$ and $78.35\%$ power reduction in FP8-INT4 and FP8-FP8 modes, respectively. Compared with FIGLUT, MxGLUT incurs a $31.95\%$ area overhead and a marginal $8.23\%$ power overhead in FP8-INT4 mode to additionally support FP8-FP8 GEMM, demonstrating an attractive trade-off between efficiency and mixed-precision functional coverage.

\subsubsection{Accelerator-Level Evaluation}
We next lift the above PE-level comparison to the accelerator level by instantiating the same four design choices as complete accelerators. Specifically, the compared systems differ only in the PE datapath, while sharing the same $64\times64$ array size, on-chip SRAM budget, VPU configuration, AXI interconnect, and DMA engine. 
As shown in Fig.~\ref{fig:mult}(c), MxGLUT achieves an area efficiency of $0.492\,\mathrm{TFLOPS/mm^2}$ and an energy efficiency of $11.58\,\mathrm{TFLOPS/W}$, representing up to $1.22\times$ and $1.65\times$ improvements over the baseline designs, respectively. 
Compared with the FP8-INT4-only FIGLUT baseline, adding native FP8-FP8 support causes only a $2.57\%$ reduction in area efficiency and a $3.34\%$ reduction in energy efficiency at the accelerator level.
These results demonstrate that MxGLUT achieves unified mixed-precision execution with negligible system-level overhead. 

\begin{figure*}[!t]
 \centering
 \includegraphics[width=0.99\linewidth]{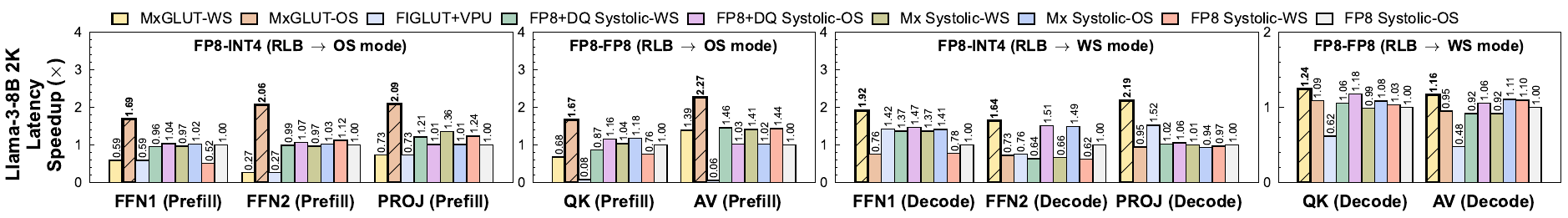}\\
 \includegraphics[width=0.99\linewidth]{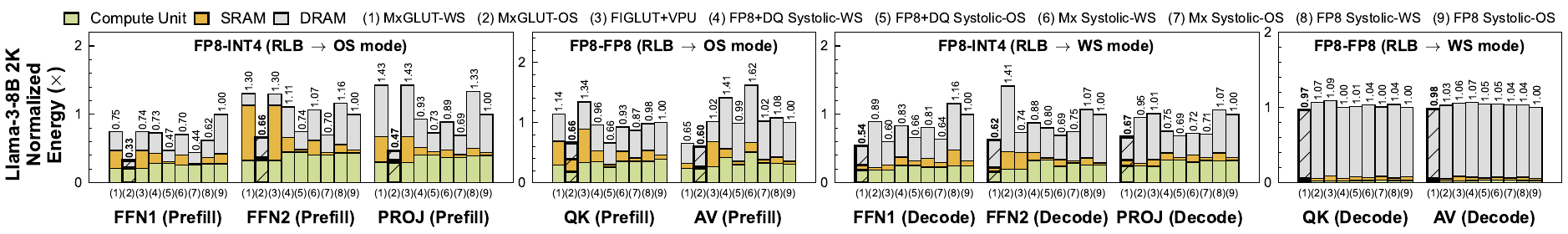}
 \caption{Layer-wise comparison of latency speedup and normalized energy across different dataflows for FP8-INT4 and FP8-FP8 GEMMs during the prefill and decode phases of Llama-3-8B.}
 \label{fig:gemm_perform}
\end{figure*}

\subsection{Performance and Energy Evaluation}\label{sec:perf}
We evaluate MxGLUT on the Llama model family in terms of latency and normalized energy under different context lengths, using prefill prompt lengths and decode KV-cache lengths of 1024, 2048, 4096, and 8192 tokens. A batch size of 64 is used to improve array utilization and enhance data reuse in GEMM execution.
We compare two configurations of the proposed accelerator, namely MxGLUT-WS and MxGLUT-OS, which correspond to the WS and OS modes of the proposed RLB dataflow. The baselines include the FP8+DQ accelerator under WS and OS dataflows~\cite{genc2021gemmini} (FP8+DQ Systolic-WS/OS), the mixed-precision systolic accelerator under WS and OS dataflows~\cite{genc2021gemmini} (Mx Systolic-WS/OS), the FP8 accelerator under WS and OS dataflows~\cite{genc2021gemmini} (FP8 Systolic-WS/OS), and FIGLUT+VPU. For the VPU-based baseline, the parallelism is fixed to 64~\cite{wang2025desa} for both nonlinear operators and GEMM, because vector-style GEMM provides limited data reuse and would otherwise require unsustainable on-chip bandwidth to sustain higher parallelism. FIGLUT+VPU is decomposed into a LUT-based FP8-INT4 path, which follows the same WS execution style as FIGLUT, and a VPU-based FP8-FP8 attention path. 

Fig.~\ref{fig:gemm_perform} first isolates the layer-wise effect of dataflow selection on Llama-3-8B at the 2K-token setting using FP8 Systolic-OS as the normalization baseline, and reveals a phase-dependent trend across the prefill and decode stages.
During prefill, where execution is dominated by large compute-bound GEMMs, MxGLUT-OS consistently achieves the best performance and energy efficiency for both FP8-INT4 linear layers and FP8-FP8 attention layers by keeping FP32 partial sums local and eliminating repeated intermediate SRAM writebacks. 
In FP8-INT4 linear layers (FFN1, FFN2, and PROJ), MxGLUT-OS achieves up to $2.09\times$ latency speedup and a normalized energy of $0.33\times$; in FP8-FP8 attention GEMMs (AV and QK), it achieves up to $2.27\times$ latency speedup and a normalized energy of $0.60\times$. The reported maximum values are computed over individual layer measurements.
These gains reflect the limitations of prior execution styles: VPU-based attention suffers from poor operand reuse, while WS execution on a LUT-oriented outer-product fabric still incurs excessive partial-sum movement during prefill. 

During decode, the preferred mode switches to WS as execution becomes increasingly memory-bound and benefits from improved weight residency, with the advantage mainly observed in FP8-INT4 linear layers. MxGLUT-WS delivers up to $2.19\times$ latency speedup and reduces normalized energy to $0.54\times$ in these layers, while FP8-FP8 attention layers show more modest gains, with up to $1.24\times$ speedup and normalized energy reduced to $0.97\times$.
This reduced gain is expected because QK and AV operations in decode are performed for a single newly generated token per request, resulting in GEMV-like execution with limited reuse along the token dimension. Therefore, although WS still improves local operand residency relative to OS, its benefit for FP8-FP8 attention is inherently constrained by off-chip DRAM traffic, which accounts for more than $93.93\%$ of the total energy in the attention layer. These layer-wise results indicate that RLB does not rely on a universally optimal dataflow; instead, it selects OS for partial-sum-dominated prefill and WS for decode workloads where weight residency provides the most benefit, especially in FP8-INT4 linear layers.

\begin{figure*}[!t]
 \centering
 \includegraphics[width=0.99\linewidth]{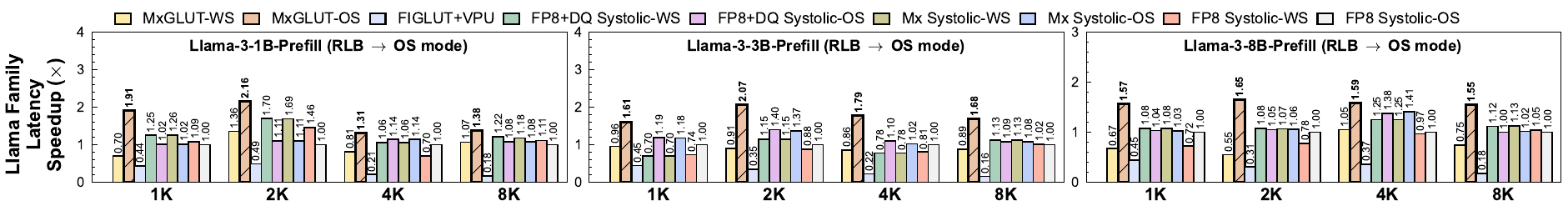}\\
 \includegraphics[width=0.99\linewidth]{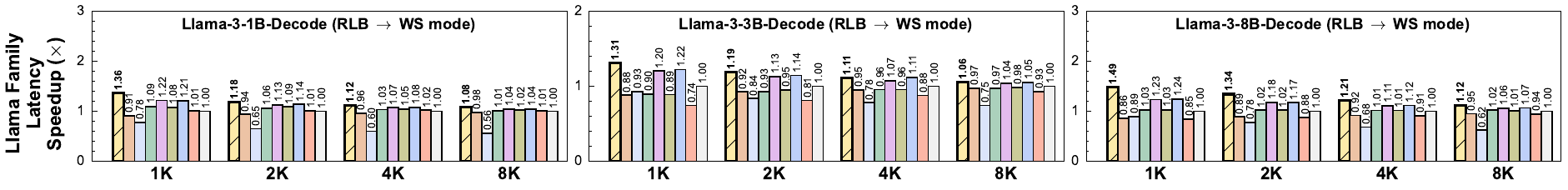}\\
 \includegraphics[width=0.99\linewidth]{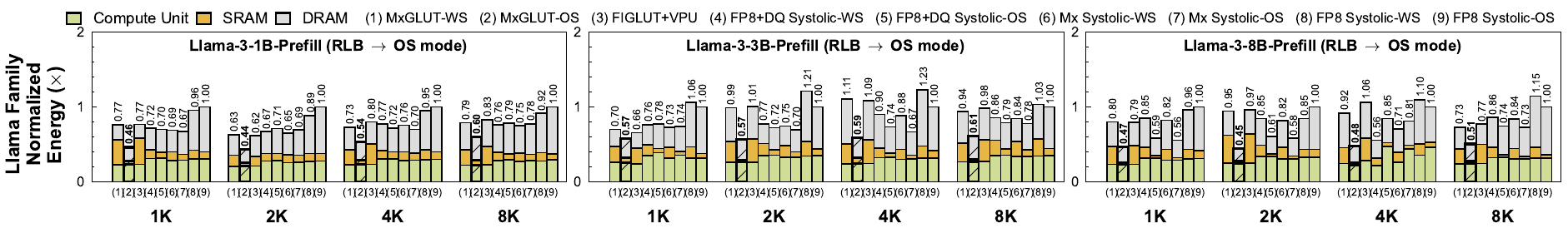}
 \includegraphics[width=0.99\linewidth]{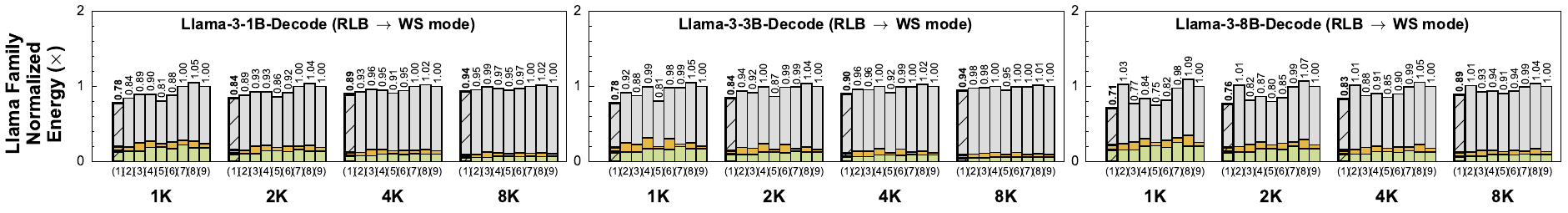}
 \caption{End-to-end inference latency speedup and normalized energy across different dataflow configurations for the Llama family during the prefill and decode phases, covering FP8-INT4 linear layers and FP8-FP8 attention GEMMs.}
 \label{fig:end2end_perform}
\end{figure*}

Fig.~\ref{fig:end2end_perform} further confirms the phase-dependent behavior across the Llama family models from 1B to 8B and across context lengths from 1K to 8K. 
During prefill, RLB selects the OS mode, which keeps FP32 partial sums local and avoids repeated intermediate writebacks. As a result, MxGLUT-OS consistently provides the best end-to-end efficiency, achieving up to $2.16\times$ latency speedup and reducing normalized energy to $0.44\times$ over FP8 Systolic-OS. 
Across the three models, the latency benefit generally reaches its maximum around the 2K context length and then decreases or saturates as attention-related DRAM traffic becomes increasingly important. 
The energy breakdown explains this trend: at moderate context lengths, prefill execution still retains a significant compute-energy component, whereas at longer context lengths, execution shifts toward DRAM-dominated behavior. For example, in Llama-3-1B prefill, the compute-unit energy share of MxGLUT-OS decreases from $50.65\%$ at 1K to $37.35\%$ at 8K.

During the decode stage, RLB selects WS to improve on-chip weight residency in GEMV-like execution. Across the Llama family, MxGLUT-WS remains the best overall configuration, achieving up to $1.49\times$ latency speedup and reducing normalized energy to $0.71\times$. 
However, the benefit decreases as context length increases, because decode becomes progressively dominated by off-chip memory traffic associated with the expanding KV-cache and weight accesses. For example, in Llama-3-3B decode at 8K, DRAM accounts for $90.08\%$ of the MxGLUT-WS energy in the end-to-end decode setting. 
This explains why WS provides clear benefits for decode linear layers but only limited additional improvement for FP8-FP8 attention at long contexts. 

The comparison among FIGLUT+VPU, MxGLUT-WS, and MxGLUT-OS in Fig.~\ref{fig:end2end_perform} further helps disentangle the contribution of native FP8-FP8 LUT execution from that of RLB dataflow selection. Compared with FIGLUT+VPU, which offloads FP8-FP8 attention GEMMs to the VPU, MxGLUT-WS achieves up to $5.94\times$ latency speedup and up to $1.16\times$ energy reduction across all evaluated model, context-length, and inference-phase settings. The latency advantage increases with context length in both prefill and decode, as attention-related FP8-FP8 computation becomes more prominent. 
In contrast, the energy gain is more modest because long-context execution is increasingly dominated by DRAM accesses.
Meanwhile, selecting the preferred RLB mode over the non-preferred MxGLUT mode provides up to $3.00\times$ latency speedup and up to $2.10\times$ energy reduction, showing that RLB contributes additional gains by adapting the stationary operand to phase-dependent execution behavior.
Together, these results show that the two components of MxGLUT are complementary and necessary for the reported end-to-end gains: native FP8-FP8 LUT execution removes the VPU bottleneck in attention, while RLB dataflow switching selects the stationary operand according to phase-dependent memory and compute bottlenecks.

\begin{table}[!t]
    \centering
    \caption{Perplexity and subnormal activation ratio of Llama models under FP8-INT4 weight-only quantization.}
    \label{tab:perplexity}
    \begin{threeparttable}
    \begin{tabular}{cccc}
        \toprule
        \textbf{Llama-3} & \textbf{1B} & \textbf{3B} & \textbf{8B} \\
        \midrule
        GPU                 & 13.54 & 10.94 & 7.14 \\
        GPU + FTZ           & 13.55 & 10.94 & 7.14 \\
        MxGLUT              & 13.77 & 11.07 & 7.20 \\
        SAR $\S$             & 0.24\% & 0.23\% & 0.33\% \\
        \bottomrule
    \end{tabular}
    \end{threeparttable}
        \begin{threeparttable}
        \begin{tablenotes}    
        \footnotesize
        \item[$\S$] Subnormal activation ratio (SAR) is defined as the fraction of FP8 activations whose exponent field is zero and mantissa is non-zero before applying FTZ normalization.
      \end{tablenotes} 
    \end{threeparttable}
\end{table}

\subsection{Accuracy Evaluation}\label{sec:ppl}
To evaluate the numerical fidelity of MxGLUT, we compare its inference perplexity with NVIDIA GPU baselines on the Llama model family using the WikiText-103 dataset~\cite{merity2016pointer}. Specifically, we consider two GPU baselines under the same FP8-INT4 weight-only quantization scheme~\cite{you2024shiftaddllm}: a standard configuration and an FTZ-enabled variant that flushes FP8 subnormal values to zero for numerical consistency. MxGLUT adopts the same quantization setting and performs FP32 accumulation to preserve partial-sum accuracy~\cite{tambe202216, 10067817}.

As reported in Table~\ref{tab:perplexity}, MxGLUT achieves comparable perplexity to the GPU baseline across all model sizes, with a maximum increase of only $1.70\%$ on Llama-3-1B, indicating negligible end-to-end accuracy degradation.
The accuracy impact mainly comes from two numerical factors: FTZ handling of FP8 subnormal values and the compact FP8-FP8 LUT-based product approximation. For FTZ, its effect is limited because the subnormal activation ratio remains consistently low across all evaluated models (SAR $\leq 0.33\%$). This is also confirmed by the GPU+FTZ results, which introduce less than $0.08\%$ perplexity variation compared with the standard GPU baseline. 
In contrast, the FP8-FP8 LUT-based product approximation introduces a larger perturbation by replacing explicit FP multiplication with a normalized and rounded 8-bit LUT representation. 
Nevertheless, its effect is already captured by the end-to-end perplexity results and remains well controlled.

\begin{figure}[!t]
    \centering
    \includegraphics[width=0.95\linewidth]{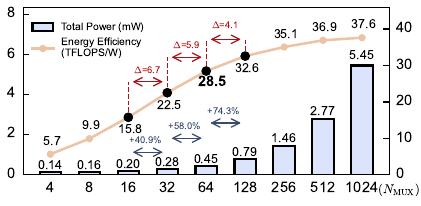}
    \caption{Design space exploration of the multiplexer count (\(N_{\mathrm{MUX}}\)) in terms of total power and energy efficiency.}
    \label{fig:dse}
\end{figure}

\subsection{Precision and Performance Scalability}

\subsubsection{Precision Scalability}
MxGLUT is not restricted to the FP8-INT4 setting adopted in the main experiments. Based on its bit-plane-oriented computation flow, the architecture can be naturally extended to support FP8-INT$x$ GEMM with x=1$\sim$16 by scaling the bit-plane processing depth. Moreover, it can also support standard FP8-FP8 GEMM in both E4M3 and E5M2 formats. These features make MxGLUT a unified computation substrate for FP8-centric weight-only mixed-precision execution and standard low-precision FP computation. Notably, MxGLUT is optimized for FP8-centric inference and does not target higher-precision formats such as FP16 or BF16, whose wider mantissas and exponent ranges would substantially increase LUT construction complexity and storage overhead.

\subsubsection{PE-Level Design Space Exploration}
We further explore the compute organization of a single MxLPE by sweeping the number of MUX ($N_{\mathrm{MUX}}$), which controls the intra-PE parallelism of delivering LUT-derived values to RACs. Increasing $N_{\mathrm{MUX}}$ improves internal parallelism and thus raises energy efficiency, but it also increases local routing complexity and switching activity, leading to higher power consumption.
As shown in Fig.~\ref{fig:dse}, increasing  $N_{\mathrm{MUX}}$ provides substantial energy-efficiency gains in the low-to-medium parallelism range. Specifically, the energy efficiency increases from $15.8\,\mathrm{TFLOPS/W}$ with 16 MUXes to $22.5\,\mathrm{TFLOPS/W}$ with 32 MUXes, and further to $28.5\,\mathrm{TFLOPS/W}$ with 64 MUXes, while the corresponding total power rises from $0.20~\mathrm{mW}$ to $0.28~\mathrm{mW}$ and $0.45~\mathrm{mW}$, respectively. Beyond the 64-MUX configuration, however, the incremental efficiency gain diminishes substantially, whereas the power overhead grows rapidly. Increasing $N_{\mathrm{MUX}}$ from 64 to 128 raises the total power by $74.3\%$, but improves the energy efficiency by only $4.1\,\mathrm{TFLOPS/W}$. A similar diminishing-return trend is observed when scaling from 128 to 256 MUXes, where power increases by a further $85.7\%$, while the energy-efficiency gain decreases to only $2.5\,\mathrm{TFLOPS/W}$.
These results identify the 64-MUX configuration as a favorable knee point in the PE design space, capturing most of the attainable energy-efficiency improvement while avoiding the disproportionate power growth incurred by larger configurations. We therefore use this configuration throughout the array-level evaluation.

\begin{table*}[!t]
\begin{center}
\caption{Comparison with representative transformer accelerators.} \label{tab:physical}
    \centering
    \begin{threeparttable}
    \resizebox{\textwidth}{!}{
    \begin{tabular}{|>{\columncolor{gray!20}}c|c|c|c|c|c|c|c|}
    \hline \rowcolor{gray!20}
    & \textbf{JSSC'22 \cite{tambe202216}} & \textbf{ISSCC'22 \cite{wang202228nm}} & \textbf{ISSCC'23 \cite{10067817}} & \textbf{ISSCC'24 \cite{kim202420}} & \textbf{VLSI'24 \cite{10631541}} & \textbf{ISSCC'25 \cite{moon2025t}} & \textbf{This Work} \\ \hline
    
    \textbf{Technology} & 16 nm & 28 nm & 12 nm & 28 nm & 22 nm & 16 nm & 28 nm \\ \hline
    
    \textbf{Model} & Transformers & \multirow{2}{*}{Transformers} & \multirow{2}{*}{Transformers} & Transformers & \multirow{2}{*}{Transformers} & \multirow{2}{*}{Transformers} & \multirow{2}{*}{Transformers} \\
    \textbf{Architecture} & RNN &  &  & (DNN, Spiking) &  &  &  \\ \hline
    
    & \multirow{2}{*}{FP8-FP8} & \multirow{2}{*}{INT12} & FP4-FP4 & \multirow{2}{*}{INT8} & FP8-FP8 & \multirow{2}{*}{INT4/8/16} & FP8-INT4 \\ 
    \multirow{-2}{*}{\textbf{Data Types}} &  &  & FP8-FP8 &  & BF16-BF16 &  & FP8-FP8 \\ \hline
    
    
    \textbf{Voltage (V)} & 0.55-1.0 & 0.56-1.1 & 0.62-1.0 & 0.7-1.1 & 0.6-1.0 & 0.45-0.85 & 0.9 \\ \hline
    
    \textbf{Frequency (MHz)} & 130-573 & 50-510 & 77-717 & 50-200 & 115-495 & 60-450 & 200 \\ \hline
    
    \textbf{Area (mm$^{2}$)} & 8.84 & 6.82 & 4.60 & 20.25 & 6.4 & 10.15 & \textbf{3.33} \\ \hline
    
    & \multirow{2}{*}{34-453} & \multirow{2}{*}{12-273} & 9-111 (FP4) & \multirow{2}{*}{22.9-47.8} & \multirow{2}{*}{49.2-461.2} & \multirow{2}{*}{7.12-152.5} & \multirow{2}{*}{141} \\
    \multirow{-2}{*}{\textbf{Power (mW)}} &  &  & 10-122 (FP8) &  &  &  &  \\ \hline
    
    \textbf{Performance} & \multirow{2}{*}{1.17} & \multirow{2}{*}{0.522} & 0.734 (FP4) & \multirow{2}{*}{3.41} & 1.01 (FP8)$\ddag$ & 0.81-2.15 (INT8) & \multirow{2}{*}{1.638} \\
    \textbf{(TFLOPS)} &  &  & 0.367 (FP8) &  & 0.51 (BF16)$\ddag$ & 0.20-0.54 (INT16) &  \\ \hline
    
    \textbf{Energy Efficiency} & \multirow{2}{*}{7.8$^{\dagger}$} & \multirow{2}{*}{4.25$^{\dagger}$} & 18.1$^{*}$ (FP4) & \multirow{2}{*}{33.4} & 3.71 (FP8)$\ddag$ & 15.2-40.3 (INT8) & \multirow{2}{*}{11.58} \\
    \textbf{(TFLOPS/W)} &  &  & 8.24$^{*}$ (FP8) &  & 2.39 (BF16)$\ddag$ & 3.8-10.1 (INT16) &  \\ \hline
    
    \textbf{Area Efficiency} & \multirow{2}{*}{0.132} & \multirow{2}{*}{0.077} & 0.159 (FP4) & \multirow{2}{*}{0.168} & 0.158 (FP8)$\ddag$ & 0.08-0.21 (INT8) & \multirow{2}{*}{\textbf{0.492}} \\
    \textbf{(TFLOPS/mm$^2$)} &  &  & 0.080 (FP8) &  & 0.080 (BF16)$\ddag$ & 0.020-0.053 (INT16) &  \\ \hline
    
    \end{tabular}
    }
    \end{threeparttable}
    \begin{threeparttable}
    \begin{tablenotes}    
        \footnotesize
        $^{\dagger}$ Measured while performing matrix multiplications. $^{*}$ Measured while performing matrix multiplications (core operations in Transformers) with 50\% input sparsity. $\ddag$ Reported feed-forward efficiency~\cite{10631541, fan202522nm}.
      \end{tablenotes} 
    \end{threeparttable}  
\end{center}
\end{table*}

\subsection{Comparison with Representative Transformer Accelerators}
Table~\ref{tab:physical} compares MxGLUT with 
representative state-of-the-art transformer accelerators 
in terms of physical overheads and efficiency. Synthesized in UMC $28\,\mathrm{nm}$ and operating at $0.9~\mathrm{V}$ and $200~\mathrm{MHz}$, 
MxGLUT delivers a peak performance of $1.638~\mathrm{TFLOPS}$ and 
supports mixed-precision FP8-INT4 and FP8-FP8 for 
accelerating weight-only quantized LLMs.

MxGLUT achieves the smallest reported area of $3.33~\mathrm{mm^2}$ among all compared designs, enabled by the elimination of dedicated FP multipliers and the unified LUT-based compute fabric. Its area efficiency of $0.492~\mathrm{TFLOPS/mm^2}$ and energy efficiency of $11.58~\mathrm{TFLOPS/W}$ represent improvements of $6.15\times$ and $1.41\times$, respectively, over the most comparable mixed-precision design (ISSCC'23~\cite{10067817}), which supports FP4-FP4 and FP8-FP8 execution. More importantly, MxGLUT is implemented in $28\,\mathrm{nm}$, which is one of the least advanced process nodes among the compared designs in Table~\ref{tab:physical}. Even under this less favorable technology setting, it still achieves the highest area efficiency and competitive energy efficiency, indicating that the observed gains primarily arise from the proposed unified LUT-based compute architecture rather than from process-node advantages.

\section{Conclusion}\label{sec:conclusion}
This paper presents MxGLUT, an RLB dataflow accelerator for efficient mixed-precision LLM inference via lookup tables. MxGLUT unifies FP8-INT4 and FP8-FP8 GEMMs within a single LUT-based compute fabric, removing dedicated FP multipliers and eliminating VPU fallback for attention GEMMs. It further introduces the RLB dataflow that switches between OS and WS mappings to match the distinct bottlenecks of prefill and decode.
Implemented in UMC $28\,\mathrm{nm}$ CMOS at $200~\mathrm{MHz}$, MxGLUT reduces multiplier area by up to $56.92\%$ and power by up to $77.07\%$ and $78.35\%$ in FP8-INT4 and FP8-FP8 modes, respectively. At the accelerator level, it achieves $0.492\,\mathrm{TFLOPS/mm^2}$ area efficiency and $11.58\,\mathrm{TFLOPS/W}$ energy efficiency, while extending native support from FP8-INT4 to unified FP8-INT4/FP8-FP8 execution incurs only $2.57\%$ and $3.34\%$ reductions in area and energy efficiency relative to the FP8-INT4-only FIGLUT baseline. 
Across the Llama family, MxGLUT achieves up to $2.16\times$ and $1.49\times$ latency speedup, and reduces normalized energy to $0.44\times$ and $0.71\times$ in prefill and decode, respectively, without significant accuracy degradation.

\section{Acknowledgement}\label{sec:acknowledgement}
The authors declare that generative AI tools, DeepSeek and Qwen, were used solely to assist with grammar correction and to improve the readability and clarity of the manuscript during its preparation. No original ideas, scientific content, analyses, results, or conclusions were generated by these AI tools. 

\renewcommand{\baselinestretch}{0.95}
\bibliographystyle{IEEEtran}
\bibliography{ref}
\vfill
\end{document}